# Decouple Electronic and Phononic Transport in Nanotwinned Structure: A New Strategy for Enhancing the Figure-of-merit of Thermoelectrics

Yanguang Zhou,[1] Xiaojing Gong,[2] Ben Xu,[3] and Ming Hu[1,4,*]

[1] *Aachen Institute for Advanced Study in Computational Engineering Science (AICES), RWTH Aachen University, 52062 Aachen, Germany*

[2] *Suzhou Institute of Nano-tech and Nano-bionics, Chinese Academy of Sciences, 215123 Suzhou, China*

[3] *School of Materials Science and Engineering, State Key Lab of New Ceramics and Fine Processing, Tsinghua University, Beijing 100084, China*

[4] *Institute of Mineral Engineering, Division of Materials Science and Engineering, Faculty of Georesources and Materials Engineering, RWTH Aachen University, 52064 Aachen, Germany*

## Abstract

Thermoelectrics (TE) materials manifest themselves in direct conversion of temperature differences to electric power and vice versa. Despite remarkable advances have been achieved in the past decades for various TE systems, the energy conversion efficiencies of TE devices, which is characterized by a dimensionless figure-of-merit ($ZT = S^2 \sigma T / \kappa_{el} + \kappa_{ph}$), remain a generally poor factor that severely limits their competitiveness and range of employment. The bottleneck for substantially boosting *ZT* coefficient lies in the strong interdependence of the physical parameters involved in electronic (*S* and *σ*, and $\kappa_{el}$) and phononic ($\kappa_{ph}$) transport. Here, we propose a new strategy of incorporating nanotwinned structures to decouple the electronic and phononic transport. Combining the new concept of nanotwin with the previously widely used nanocrystalline approach, the power factor of the Si nanotwin-nanocrystalline heterostructures is enhanced by 120% compared to bulk crystalline Si, while the lattice thermal conductivity is reduced to a level well below the amorphous limit, yielding a theoretical limit of 0.43 for *ZT* coefficient at room temperature. This value is almost two orders of magnitude larger than that for bulk Si and twice of the polycrystalline Si. Even for the experimentally existing nanotwin-nanocrystalline heterostructures (e.g. grain size of 5 nm), the *ZT* coefficient can be as high as 0.2 at room temperature, which is the highest *ZT* value among all the Si based bulk nanostructures so far. Such substantial improvement stems from two aspects: (1) the improvement of the power factor is caused by the increase of Seebeck coefficient (degeneracy of the band valley) and the enhancement of electrical conductivity (the reduction of the effective band mass); (2) the significant reduction of the lattice thermal conductivity is mainly caused by the extremely strong phonon-grain boundary and phonon-twin boundary scattering. Our results suggest that nanotwin is an excellent building-block for enhancing the TE performance in diamond-like semiconductors and our study provides a new strategy to the innovative development of other TE materials.

---

[*] Author to whom all correspondence should be addressed. E-Mail: hum@ghi.rwth-aachen.de (M.H.)

# 1. Introduction

Thermoelectric (TE) devices enable solid-state cooling to replace cumbersome vapor-compression cycle technologies as well as electricity generation from a variety of waste heat sources such as industries and vehicles [1-4]. The performance of the TE materials is characterized by a dimensionless figure-of-merit (*ZT*) which can be written as $ZT = S^2\sigma T/(\kappa_{el} + \kappa_{ph})$, where $S$, $\sigma$, $\kappa_{el}$ and $\kappa_{ph}$ are the thermoelectric power (Seebeck coefficient), electrical conductivity, electronic thermal conductivity and lattice (phonon) thermal conductivity, respectively. An ideal TE should be electrical conductor [high power factor (PF) $S^2\sigma$], but thermal insulator [low thermal conductivity $(\kappa_{electron} + \kappa_{phonon})$]. This conflicting requirement poses a material challenge, due to the strong coupling effect between the electronic and phononic transport. In general, development schemes to improve TE conversion efficiency in the past decades were guided by the concept of "phonon glass – electron crystal" [5-7], i.e., reducing the lattice contribution to the thermal conductivity as closely as possible to an amorphous state, while keeping relatively high electrical conductivity and Seebeck coefficient [8]. Nanostructuring the existing bulk TE materials has been proved to be a quite efficient way to obtain decent thermoelectric performance [1, 4, 9-17], due to the faster reduction of the lattice thermal conductivity than the slower decrease of PF, which means the thermal conductivity could approach the lowest possible value, while the PF of nanostructured materials might not be the largest yet (at least currently it is generally smaller than that of the bulk counterpart). Therefore, there is still plenty of room to improve the *ZT* value further via nanostructuring.

Recently, experimental results [18] show that a special grain boundary in materials, the twin boundary (TB), can enhance the current transport properties in cubic CdTe. Considering the similarity among all the diamond-like structures and experimental feasibility of generating TB in such structures [18-21], an idea is inspired: by introducing TB into diamond-like semiconductors such as CdTe, Si, Ge, SiC, BN and BeO, the PF could be maintained or even increase. Since it is easy to know the phonon transport can be hindered by the TB due to the phonon-twin boundary scattering, the electronic and phononic transport could be decoupled (the so-called bottleneck problem for thermoelectrics) and thus can be modulated in opposite direction, if the twin boundary is used in thermoelectrics.

In this Letter, by combining first-principle calculation, Boltzmann transport equation (BTE) and classical molecular dynamics (MD) simulations, we perform a proof of concept computer experiment and provide evidence that the PF can be improved with respect to its bulk counterpart and the thermal conductivity can be decreased simultaneously by TB in the diamond-like semiconductors. Furthermore, by combining polycrystalline (normal grain boundary) and TB structures, we show that the thermal conductivity can be reduced down to extremely low values, eventually much lower than the amorphous limit, while the PF can be maintained to be at the same level or even enhanced compared to their counterpart. As a result, for the limit grain size (around 1 nm) that is achievable in experiment [22], the *ZT* of Si can reach as large as 0.43 which is almost two orders of magnitude higher than that of bulk crystalline Si and twice of the polycrystalline Si.

## 2. Results and Discussion

Our model structure consists of perfect $\Sigma 3$ TBs in Si which has coherent and mirror plane [(111) plane] symmetry, as shown in Fig. 1(a). Si is chosen as the model system because the nanotwinned structures have been widely fabricated in experiments [21, 23, 24]. The atoms on the TB [red atoms in Fig. 1(a)] are stacking in the hexagonal crystal structure and the bulk atoms [yellow atoms in Fig. 1(a)] have the diamond crystal structure. Meanwhile, in order to explore the electronic structure of TB, the electronic density isosurface of TB is shown in Fig. 1(b). The charge distribution on the TB (circle A in Figure 1b) has been strongly altered with respect to a pristine diamond structure (circle B in Figure 1b), a feature which is likely expected to affect electronic transport. Four bulk Si containing TB (hereafter denoted as Si-TB) structures with different periodic length ($\lambda_{TB}$) are considered (Figure 1a), namely Si-TB-1(2, 3, 4) with $\lambda_{TB}$ of 1.9(3.8, 5.7, 9.6) nm, respectively. All band structures of Si-TB and perfect crystalline Si are calculated using Vienna Ab-initio Simulation Package (VASP) based on the density functional theory (DFT) [25]. Periodic boundary conditions are applied in all three directions. The Si atoms are described by the pseudopotential containing $s^2p^2$ as valence states with local-density approximation (LDA) in the Project-Augmented-Wave (PAW) exchange-correlation function. A plane-wave basis is implemented with kinetic energy cutoff of 320 eV.

At first, the lattice constant and geometry of the systems are optimized until the energy difference and the Hellman-Feynman force are converged within $1 \times 10^{-4}$ eV and $5 \times 10^{-3}$ eV/Å, respectively, using a $4\times4\times4$ to $10\times10\times10$ shifted Monkhorst-Pack mesh of **K** points to sample the first Brillouin zone. Then, the electronic transport properties of the bulk crystalline Si and Si-TBs are calculated by the BoltzTrap [26] based on the semi-classical BTE. A quite dense **K** points mesh, $15\times15\times15$, is employed to ensure the accuracy of the Kohn-Sham eigenvalues which are used as input in the BoltzTrap. It is worth noting that the band gap computed from Kohn-Sham states is systematically underestimated [27], but it does not affect directly the comparison and conclusion since all the calculations are performed with the same method, which has been proved by previous studies [26, 28, 29]. All the results of $\kappa_{ph}$ are calculated by Green-Kubo equilibrium molecular dynamics (GK-EMD) simulations [30-32] (details can be found in Sec. 4), in which the interatomic interaction is described by Tersoff potential [33]. We would like to point out that, the Tersoff potential normally overestimates the $\kappa_{ph}$ of bulk Si: the $\kappa_{ph}$ computed by Tersoff potential is around 200 W/mK while the experimental value is 156 W/mK [34]. However, all of our results are calculated using the same methods and interatomic potential. Therefore, this overestimation is believed to have no impact on our comparison and conclusion.

## 2.1. Thermal and Electrical Properties of Nanotwin

Figs. 1(c-f) show the electronic transport properties in bulk Si containing TB as well as in perfect crystalline Si in the case of N-type doping (as been plotted in Fig. 2 below, only the conduction band is modulated by the TB structure). Obviously, the electronic transport properties of the Si-TB structure should be direction dependent since the three crystal

directions are inequivalent. Due to the degeneracy of multiple valleys and the increase of band gap (see the electronic band structure and discussion below) caused by the narrow spaced TB (Si-TB-1, Si-TB-2 and Si-TB-3), the absolute value of the Seebeck coefficient of the nanotwinned structure is unexpectedly increased with respect to bulk crystalline Si [Fig. 1(c)]. Meanwhile, it is quite interesting to find that along one of the in-plane directions ([112] direction), the electrical conductivity can be improved compared to the bulk crystalline Si due to the TB structure (detailed discussion can be found below): the smaller the $\lambda_{TB}$ of the nanotwinned structure, the larger the enhancement of the electrical conductivity [Fig. 1(d)]. Following the Wiedemann-Franz law [35], the $\kappa_{el}$ should have the similar trend with the electrical conductivity [Fig. 1(e)]. Finally, due to the combined effect of the Seebeck coefficient and electrical conductivity, for all the cases the PF along [112] direction can be improved and the case with $\lambda_{TB}$ of 1.9 nm can be even doubled with respect to the perfect crystalline Si. The PF along [110] direction with $\lambda_{TB}$ of 1.9 and 3.8 nm also increases slightly.

Now, we investigate the underlying mechanism that leads to the enhancement of Seebeck coefficient and electrical conductivity for the nanotwinned structure. It is well-known that the material parameter called the thermoelectric quality factor ($B$) determines the optimized figure-of-merit [36-40]

$$ZT \propto B = \left(\frac{k_b}{e}\right)^2 \frac{2e(k_b T)^{3/2}}{(2\pi)^{3/2} \hbar^3} \frac{N_V \mu_0 m_b^{*3/2}}{\kappa_{ph}} T, \qquad (1)$$

where $m_b^*$ is the band effective mass (details can be found in Sec. 4), $N_v$ is the band degeneracy, $\mu_0$ is the mobility at the nondegenerate limit, $k_b$ is the Boltzmann constant, $\hbar$ is the reduced Planck constant, $e$ is free electron mass. The traditional method to increase $B$ is improving Seebeck coefficient via increasing $m_b^*$ [41]

$$S = \frac{2k_b T}{3e\hbar^2} m_d^* \left(\frac{\pi}{3n_{carrier}}\right)^{2/3}, \quad (2)$$

with

$$m_d^* = N_V^{2/3} m_b^*, \quad (3)$$

in which, $n_{carrier}$ is the carrier density and $m_d^*$ is the effective mass of density of state. However, such method is detrimental to the electrical conductivity [42] since the $\mu_0$ can be written as [36, 43]

$$\mu_0 = \frac{2^{3/2} \pi^{1/2} \hbar^4 eC}{3 m_I^* \left(m_b^* k_b T\right)^{3/2} \Xi^2}, \quad (4)$$

where $m_I^*$ is the inertial effective mass and equal to $m_b^*$ of the concerned current transport direction, $C$ is the elastic constant, and $\Xi$ is the deformation potential coefficient. Combining Eqs. (1) and (4), we can know

$$ZT \propto B = \frac{2k_b^2 \hbar T}{3\pi} \frac{CN_V}{3m_I^* \Xi^2 \kappa_{ph}}. \quad (5)$$

Therefore, both low effective band mass $m_b^*$, which can guarantee high electrical conductivity [Eq. (4)], and large valley degeneracy $N_V$, which can gain large Seebeck coefficient [Eq. (2)], are quite beneficial for achieving high figure-of-merit [Eq. (5)]. From our results (Fig. 2), it is clearly seen that the valley degeneracy $N_V$ in the conduction band is effectively increased to 15 (1×Γ, 2×X, 8×W and 4×H), 9 (1×Γ, 2×A, 4×K and 2×Y), 9 (1×Γ, 2×A, 4×K and 2×Y) and 9 (1×Γ, 2×A, 4×K and 2×Y) when their band extrema are converged to have energies within a few (5) $k_bT$ of each other. For the bulk crystalline Si, the $N_V$ in the conduction band is only 6 (6×A). The large improvement of the $N_V$ in the conduction band will, in return, lead to enhancement of Seebeck coefficient [Eqs. (2) and (3)] in the Si-TB structures as shown in Fig. 1. Meanwhile, Si-TB-1 will possess the largest Seebeck coefficient where the largest $N_V$ is found. It is worth noting that the Si-TB here can be also regarded as a special superlattice. It is well-known that the superlattice structures can be used to manipulate the symmetrically inequivalent bands to obtain high degeneracy $N_V$ [44]. At the same time, the effective band mass $m_b^*$ of the conduction band minima (CBM) is calculated (details can be found in Sec. 4). It is easy to find that the $m_{b,y\,or\,z}^*$ and $m_{b,x}^*$ of the Si-TBs are much smaller and larger than that of the bulk crystalline Si (Table I), respectively. Both the $m_{b,y\,or\,z}^*$ and $m_{b,x}^*$ increase with $\lambda_{TB}$. At the same time, as shown in Fig. 2, the band gap will be enlarged by the TB structure, which means that it becomes more difficult for electrons to translate from valence band to conduction band, and then, is detrimental to the electrical conductivity. Therefore, there are two competing mechanisms responsible for the variation of the electrical conductivity of Si-TBs as shown in Fig. 2: (1) the decrease of effective band mass $m_b^*$ leads to the improvement of mobility $\mu_0$ [Eq. (4)];

(2) the increase of the band gap leads to the difficulty of thermal excitation of electrons from valence band to conduction band. Only when the former factor is dominant, the electrical conductivity can be improved such as *z*-direction ([112] direction) in the Si-TBs. For [110] direction, although the effective band mass $m_{b,y}^*$ is reduced, the enhancement of band gap will be dominant. Therefore, the electrical conductivity of Si-TBs along [110] will decrease slightly with respect to its bulk counterpart. For electrical conductivity along the [111] direction, both the increase of the $m_{b,x}^*$ and the improvement of band gap are responsible for its large reduction.

Meanwhile, there is no surprise to find that the $\kappa_{ph}$ is significantly reduced with respect to the bulk crystalline Si (results are listed in Supporting Information (SI) Section I), since the TB is actually a special kind of grain boundary and thus scatters the phonons. Finally, we can know that the *ZT* of Si-TBs can be augmented by as large as around 3 times (up to 0.01 at room temperature) compared to the bulk crystalline counterpart. Despite the large enhancement, the pure Si-TB structure is not a good thermoelectrics since the $\kappa_{ph}$ is too high.

## 2.2. Thermoelectric Performance of Si Nanotwin Nanomembranes

Next, we combine the nanotwin concept with the method of nanostructuring to further enhance the thermoelectric performance. Note that the nanocrystalline materials with TBs inside the grains with diamond-like structure have been fabricated in experiments [19]. At first, the Si twin-boundary nanomembrane (denoted as Si-TBNM) is studied [Figs. 3(e-f)]. The electrical transport properties of the Si-TBNM with grain size of 3 nm and layer thickness of about 3 nm are shown in Figs. 3(a-d) (the calculation details can be found in SI Section II). Although the electrical conductivity of Si-TBNMs is decreased compared to bulk crystalline Si, the PF of Si-TBNMs is comparable to the bulk value, due to the large enhancement of Seebeck coefficient. For the carrier density of $1\times10^{20}$ cm$^{-3}$, the PF of Si-TBNMs with TB-1, TB-2, TB-3 and TB-4 incorporated is 80%, 70%, 40% and 30% of the bulk value, respectively. Meanwhile, both the simulation results and theoretical predictions (details see SI Section III) show that the $\kappa_{ph}$ of such TBNMs can be reduced by almost two orders of magnitude with respect to the bulk value [Fig. 4(a)]. The $\kappa_{ph}$ of these Si-TBNMs with grain size smaller than around 6 nm is even lower than the amorphous limit. Such phenomenon is also reported in previous studies [30, 45, 46]. Increasing the thickness of the membrane leads to the increase of the $\kappa_{ph}$ of Si-TBNMs due to the weakened phonon-boundary scattering [Fig. 4(b)]. This is also the reason why the Si-TBNMs with larger grain size has a stronger thickness dependence: the larger the grain size, the longer mean free path of the phonons the Si-TBNMs have [Fig. 4(c)], and then, the phonon-surface scattering is more evident. From Fig. 4(c), we can also find that the largest MFP of phonons is much smaller than the grain size: for instance, the maximum MFP of phonons for the Si-TBNMs is only 3.5 nm when the grain size is around 8 nm. This indicates that the thermal conductivity of Si-TB nanostructures can break the Casimir limit which assumes the average MFP of phonons should be equal to the grain size [47]. The phonons with MFP in the range of 1.0 to 3.0 nm for the Si-TBNMs with grain size of around 8 nm contribute around 95% of total $\kappa_{ph}$.

Meanwhile, both the maximum MFP of the phonons and the range of the MFP of the dominant phonons decrease with grain size decreasing. Also, relatively larger MFP phonons can exist in the TBNMs when the layer thickness increases.

From Fig. 5(a), it is easy to find the *ZT* coefficient of the Si-TBNMs can reach 0.23 at room temperature, which is about 60 times of the bulk crystalline Si, when the grain size is around 3.0 nm. Such high *ZT* value is mainly due to the extremely low $\kappa_{ph}$ and the bulk comparable PF. With grain size or layer thickness increasing, the *ZT* coefficient will be reduced which is mainly due to the increase of $\kappa_{ph}$. To further improve the thermoelectric performance, we propose three-dimensional Si nanotwin-nanocrystalline heterostructures (NT-NC HSs) which are easier to be fabricated in experiments, and calculated the corresponding *ZT* value (see SI Sections II and III for details). It is surprising to find that the PF of the Si NT-NC HSs can be even larger than that of bulk crystalline Si [Fig. 5(b)]. Such enhancement of the PF is mainly caused by the contribution from the other in-plane direction ([110] direction). At the same time, it is easy to know that $\kappa_{ph}$ of the Si NT-NC HSs is even lower than that for the Si TBNMs with the same grain size. This is because in our models the grain boundary is rougher than the surface (SI Section III).

## 2.3. The Optimal Figure-of-merit of Si Nanotwin Nanocrystalline Heterostructures

The next question is: what is the maximum possible *ZT* value of the Si NT-NC HSs? From Figs. 5(b) and 5(c), we find both PF and $\kappa_{ph}$ are grain size dependent. This indicates that an optimal figure-of-merit of Si NT-NC HSs can be obtained by modulating the grain size. Fig. 5(c) shows that in the experimental limit of grain size [22] (around 1 nm), the *ZT* value of the Si NT-NC HSs can be as large as 0.43 at room temperature (theoretical limit), which means Si NT-NC HS could be the highest thermoelectric material among all the Si-based bulk structures reported so far, e.g. polycrystalline Si with grain size of 100 nm (*ZT*: ~0.033) [48], , bulk SiGe alloy (maximum *ZT*: 0.08) [15] and nanostructured SiGe alloy (maximum *ZT*: ~0.16) [11, 17, 49]. Even when the grain size is increased to around 5 nm, which is the case that the bulk Si NT-NC HSs has already been fabricated in experiments [23], the *ZT* value of the bulk Si NT-NC HSs (around 0.2) is still higher than most of traditional Si-based bulk thermoelectrics. It is also worth noting that the *ZT* value predicted in our paper is actually underestimated, which means the real *ZT* value of the bulk Si NT-NC HSs could be even higher in practice.

## 3. Conclusions

In conclusion, a new strategy based on lattice defect engineering is proposed to improve the thermoelectric performance of the diamond-structure semiconductors: simultaneously increasing the power factor and decreasing the thermal conductivity by introducing nanotwins. The enhancement of the power factor in nanotwinned structures is caused by two reasons: (1) the improvement of the Seebeck coefficient is due to the degeneracy of multiple valleys ($N_V$); (2) the increase of the electrical conductivity is caused by the reduction of the effective band mass ($m_b^*$). The reduction of the thermal conductivity (mainly lattice component) is attributed to the phonon-twin boundary scattering. Consequently, the figure-of-merit of the Si nanotwinned structure can be improved by 3 times with respect to bulk crystalline counterpart. Furthermore, when the nano-twins are incorporated into the bulk Si nanocrystallines, surprisingly, the power factor of the Si nanotwin-nanocrystalline heterostructures can be even larger than that of bulk crystalline Si and the lattice thermal conductivity is well below the Casimir limit. As a result, the figure-of-merit of the heterostructure can reach as large as 0.43 at room temperature, which is two orders of magnitude improvement compared to bulk crystalline Si and even 53% higher than that for nanocrystalline Si. Our results suggest that incorporating nanotwin as structure engineering is a promising strategy for breaking the bottleneck in the semiconductor thermoelectrics through decoupling the electronic and phononic transport.

## 4. Experimental Section

*Thermal Conductivity Calculation Using the EMD Simulations:* The Green-Kubo (GK) method is used to utilized to predict the lattice thermal conductivity $\kappa_{ph}$ of the NT and TBNMs, which can be calculated by integrating the time-dependent heat-flux autocorrelation functions (HFACF) through [50]

$$\boldsymbol{\kappa} = \frac{V}{k_B T^2} \int_0^\infty \langle \mathbf{J}(\tau) \cdot \mathbf{J}(0) \rangle d\tau, \quad (6)$$

where $k_B$ is Boltzmann constant, $V$ is system volume, $T$ is system temperature and $\tau$ is auto-correlation time. The angular bracket means the ensemble average. **J** is the heat flux vector and has the expression of[51]

$$\mathbf{J} = \frac{1}{V}\left[\sum_{i}^{N} e_i \mathbf{v}_i + \frac{1}{2}\sum_{i,j;i\neq j}^{N}(\mathbf{F}_{ij}\cdot\mathbf{v}_i)\mathbf{r}_{ij} + \sum_{i,j,k;i\neq j\neq k}^{N}(\mathbf{F}_{ijk}\cdot\mathbf{v}_i)(\mathbf{r}_{ij}+\mathbf{r}_{ik})\right], \quad (7)$$

in which $e_i$ is the energy of atoms, $\mathbf{v}_i$ is the velocity of atoms, $\mathbf{r}_{ij}$ is the distance between two atoms $i$ and $j$, $\mathbf{F}_{ij}$ and $\mathbf{F}_{ijk}$ are the two-body and three-body force, respectively. A timestep of 0.05 fs is used for all systems. Firstly, 500 ps run with *NPT* (constant particles, constant pressure and constant temperature) ensemble is implemented to relax the structure. Following equilibrium, 500 ps to 2 ns run is carried out with *NVE* (constant particles, constant volume without thermostat) ensemble.

*The Effective Band Mass Calculation at CBM :* In order to understand the mechanism that causes the improvement of the electrical conductivity, we calculate the effective band mass of the CBM, e.g., Γ point and A point [Fig. 2]. It is well-known that for an external electrical field the effective mass of a charger carrier is defined as

$$\left(\frac{1}{m_b^*}\right)_{ij} = \frac{1}{\hbar^2}\frac{\partial^2 E_n(\mathbf{K})}{\partial \mathrm{K}_i \mathrm{K}_j}, \quad i,j = x,y,z, \quad (8)$$

where $i$ and $j$ indicate the reciprocal components, and $E_n(\mathbf{K})$ is the dispersion relation for the *n*-th band. Meanwhile, for group III-IV semiconductors, the energy $E_n(\mathbf{K})$ of the wave vector $\mathbf{K}$ at the band minimum or maximum can be written as

$$E(\mathbf{K}) = E_0 + \frac{\hbar^2}{2m_{b,x}^*}\mathrm{K}_x^2 + \frac{\hbar^2}{2m_{b,y}^*}\mathrm{K}_y^2 + \frac{\hbar^2}{2m_{b,z}^*}\mathrm{K}_z^2. \quad (9)$$

Combining Eq. (9) and (10), we have

$$\left(\frac{1}{m_b^*}\right)_{ij} = \frac{1}{\hbar^2} \begin{pmatrix} \frac{\partial^2 E_n(\mathbf{K})}{\partial K_x^2} & \frac{\partial^2 E_n(\mathbf{K})}{\partial K_x K_y} & \frac{\partial^2 E_n(\mathbf{K})}{\partial K_x K_z} \\ \frac{\partial^2 E_n(\mathbf{K})}{\partial K_y K_x} & \frac{\partial^2 E_n(\mathbf{K})}{\partial K_y^2} & \frac{\partial^2 E_n(\mathbf{K})}{\partial K_y K_z} \\ \frac{\partial^2 E_n(\mathbf{K})}{\partial K_z K_x} & \frac{\partial^2 E_n(\mathbf{K})}{\partial K_z K_y} & \frac{\partial^2 E_n(\mathbf{K})}{\partial K_z^2} \end{pmatrix}. \qquad (10)$$

Finally, the effective mass at the conduction band minima is calculated using finite difference method [https://github.com/afonari/emc].

## Acknowledgments

M.H. gratefully acknowledges Prof. Luciano Colombo (University of Cagliari) for valuable comments and proof reading the manuscript. Simulations were performed with computing resources granted by the Jülich Aachen Research Alliance-High Performance Computing (JARA-HPC) from RWTH Aachen University under Project No. jara0155 and Special Program for Applied Research on Super Computation of the NSFC Guangdong Joint Fund (the second phase). Y.Z. would like to thank Dr. Long Cheng (RWTH Aachen University) for his helpful and fruitful discussions.

## Supporting Information

Supporting Information Available: I. The lattice thermal conductivity of the Si-TBs and Bulk Si. II. Electrical transport properties of nanostructures. III. Theory to calculate the lattice thermal conductivity of nanostructures. IV. Inherent relaxation time of electrons.

## Notes

The authors declare no competing financial interests.

Table I. The effective band mass at conduction band minima [Γ point for nanotwinned Si with different periodic length (Si-TB-1/2/3/4) and A point for bulk crystalline Si]

| Type | $m_{b,x}^*$ ($m_e$) | $m_{b,y}^*$ ($m_e$) | $m_{b,z}^*$ ($m_e$) |
|---|---|---|---|
| Bulk Crystalline Si | 0.147 | 0.147 | 0.972 |
| Si-TB-1 | 0.153 | 0.031 | 0.199 |
| Si-TB-2 | 0.171 | 0.033 | 0.247 |
| Si-TB-3 | 0.451 | 0.070 | 0.615 |
| Si-TB-4 | 0.888 | 0.300 | 0.904 |

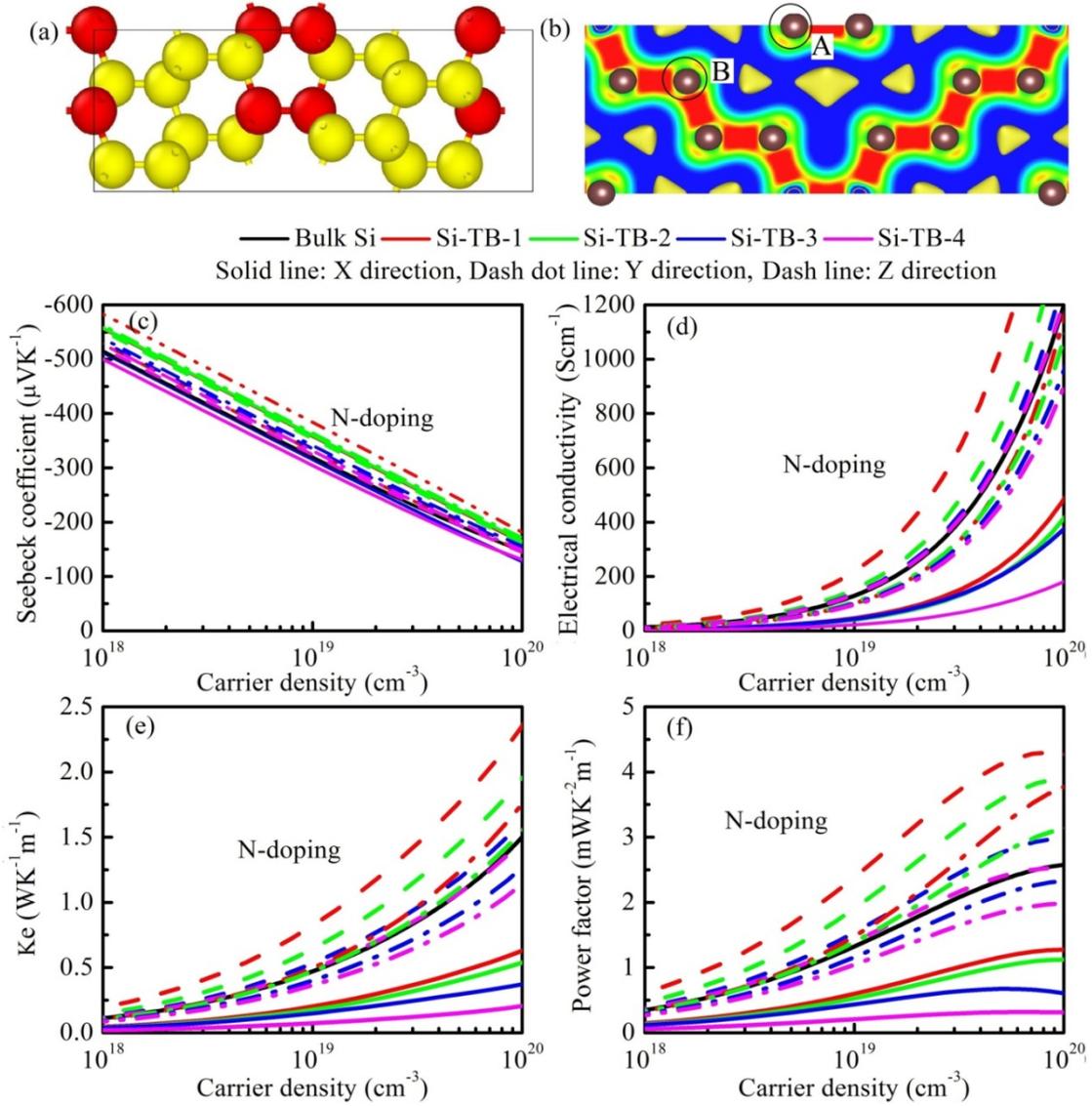

FIG. 1. (a) The schematic of a representative Si nanotwinned structure (periodic length of 1.9 nm). The atoms on the twin boundary that are stacking in the hexagonal crystal structure and the bulk atoms with the diamond crystal structure are highlighted in red and yellow, respectively. (b) The corresponding electrical charge density. The electrical transport properties of the nanotwinned structure along three directions: (c) Seebeck coefficient, (d) electrical conductivity, (e) electronic thermal conductivity and (d) power factor. The X, Y, and Z direction are [111], [110], and [112], respectively.

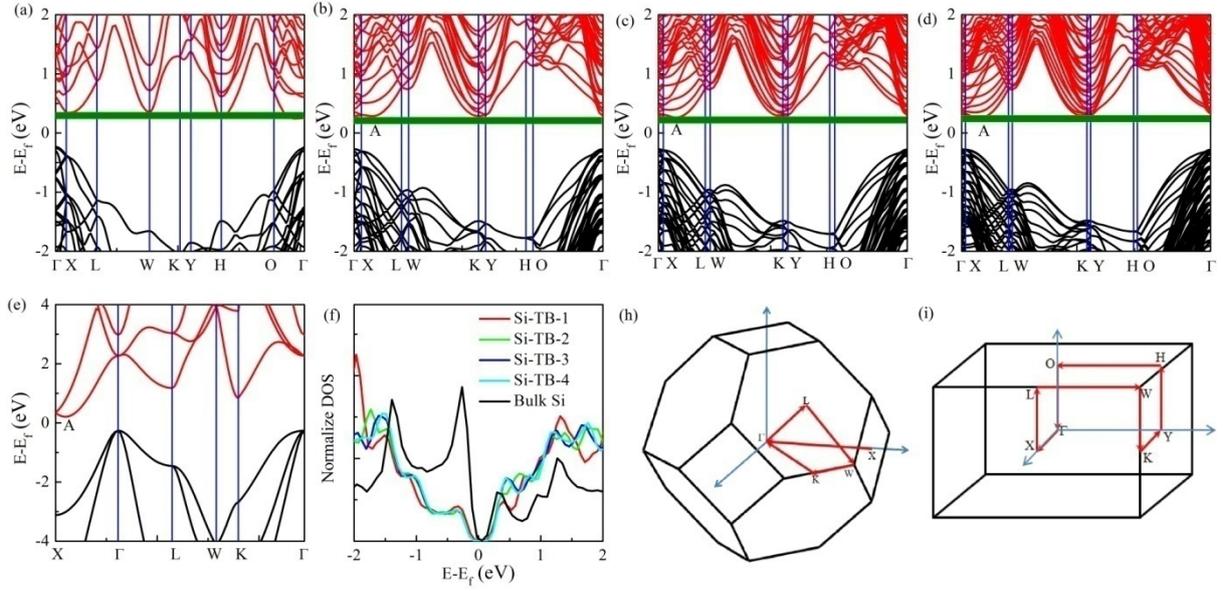

FIG. 2. Band structures of the Si nanotwinned structure with various periodic length: (a) 3.8 nm, (b) 7.6 nm, (c) 11.4 nm, and (d) 15.2 nm. The horizontal lines denote the energy level of 5 $k_b$T. (e) The results for bulk crystalline Si. (f) Comparison of the density of states between bulk crystalline Si and Si-TBs. The Brillouin zone of the bulk crystalline Si (h) and Si-TBs (i).

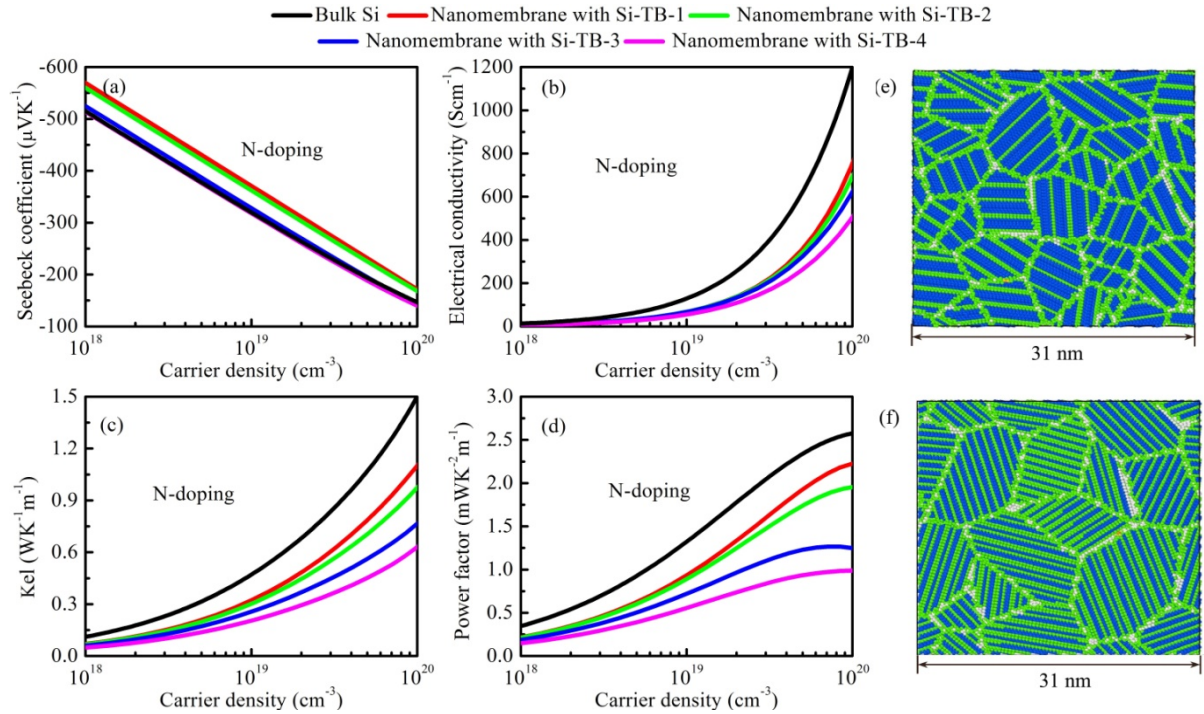

FIG. 3. Electrical transport properties of the Si nanomembranes with nanotwins inside the grains: (a) Seebeck coefficient, (b) electrical conductivity, (c) electronic thermal conductivity, and (d) power factor. (e, f): the atomic structures of the Si nanotwinned membranes with different periodic lengths (e, 3.8 nm; f, 1.9 nm). The atom is colored by the coordination number analysis.

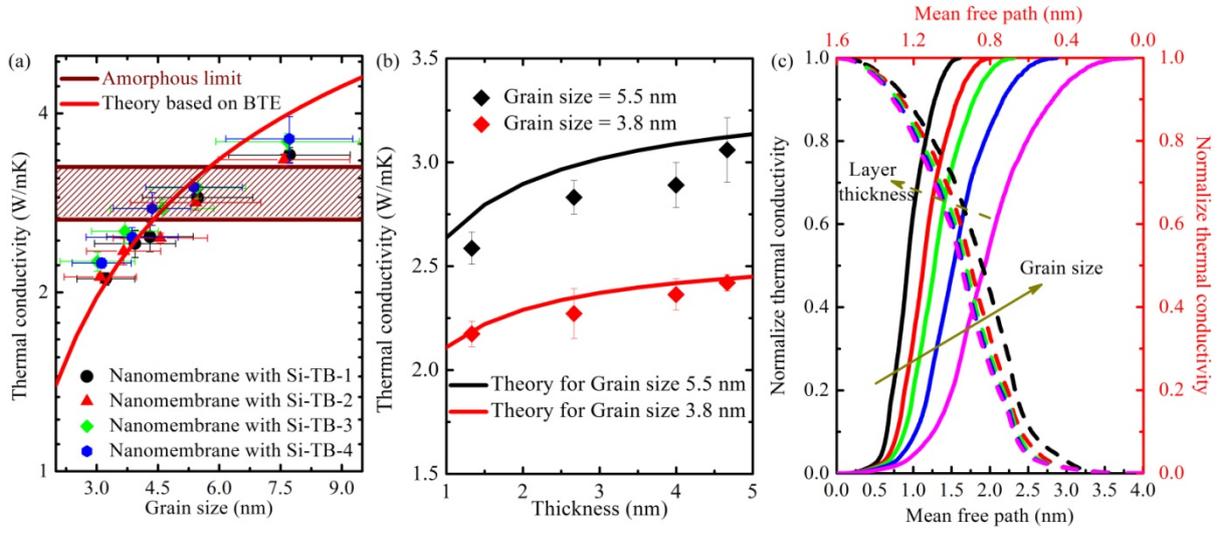

FIG. 4. Grain size (a) and thickness (b) dependent lattice thermal conductivity of the Si nanomembranes incorporated with nanotwins. The symbols are results calculated by Green-Kubo equilibrium molecular dynamics simulation and the solid lines are theoretical results. (c) The phonon mean free path vs. grain size (using the black bottom-left axes) and thickness (using the red top-right axes). The dashed and solid lines are for layer thickness and grain size dependence, respectively. The arrows denote the increase direction of layer thickness and grain size.

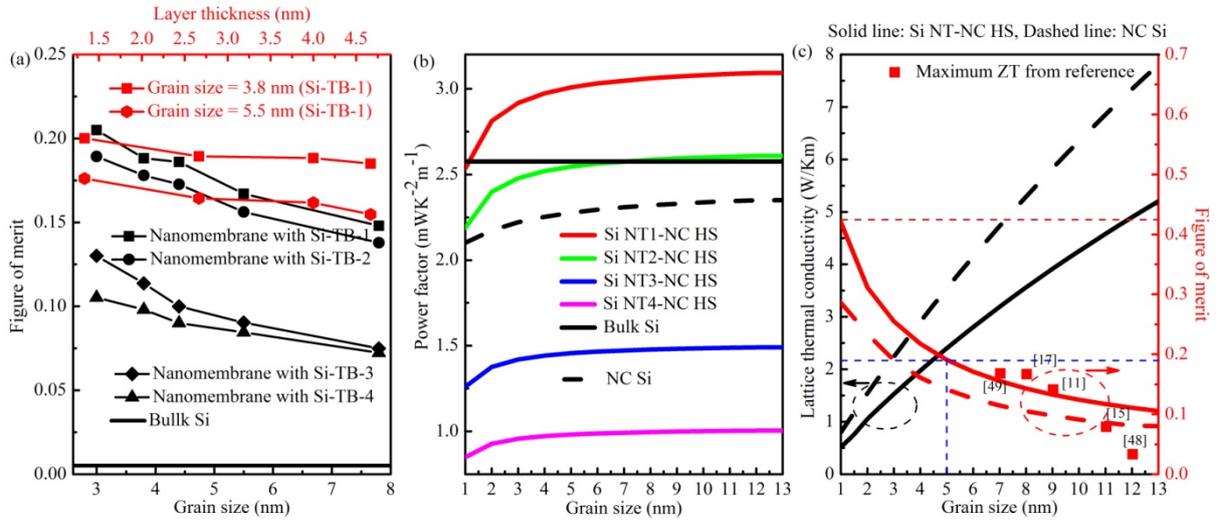

FIG. 5. (a) The figure of merit (*ZT*) of the Si-TB NMs vs. grain size (using the black axes) and layer thickness (using the red axes). (b) Power factor for the Si nanotwin-nanocrystalline heterostructures with different periodic lengths and nanocrystalline Si (NT-NC HSs or NC). The black horizontal line denotes the power factor for bulk crystalline Si. (c) Lattice thermal conductivity (the black axis) and figure of merit (the red axis) of the Si NT-NC HSs (solid lines) and nanocrystalline Si (long dashed lines) vs. grain size. The square symbols are the *ZT* coefficient for the Si-based nanostructures reported in literature with reference number labeled. The red short dashed line represents the maximum *ZT* (0.43) achieved in our proposed Si NT-NC HSs (theoretical limit). The blue short dashed line represents the *ZT* (0.2) that can be achieved in the experimentally existing Si NT-NC HSs (grain size of 5 nm).